\documentstyle[prl,aps]{revtex}

\begin{document}
\draft

\twocolumn[\columnwidth\textwidth\csname@twocolumnfalse\endcsname
 
 
\title{The nuclear quadrupole moment of $^{57}$Fe  from 
microscopic nuclear  and atomic  calculations}

\author{
Gabriel Mart\'{\i}nez-Pinedo,$^{1,2}$
Peter Schwerdtfeger,$^3$
Etienne Caurier,$^4$
Karlheinz Langanke,$^1$
Witold Nazarewicz, $^{5-7}$ and
Tilo S\"{o}hnel\,$^3$ 
}

\address{%
$^1$Institut for Fysik og Astronomi, {\AA}rhus Universitet,
DK-8000 {\AA}rhus C,
Denmark}

\address{%
$^2$Departement f\"ur Physik und Astronomie, Universit\"at Basel, Basel,
Switzerland}

\address{%
$^3$Department of Chemistry,
  University of Auckland,
  Private Bag 92019,
  Auckland,
  New Zealand}

\address{%
$^4$Institute de Recherches Subatomiques (IN2P3-CNRS-Universit\'e Louis
Pasteur), 
F-67037 Strasbourg Cedex 2,
France}

\address{%
$^5$Department of Physics, University of Tennessee, Knoxville,
             Tennessee 37996}

\address{%
$^6$Physics Division, Oak Ridge National Laboratory,
                Oak Ridge, Tennessee 37831}
\address{%
$^7$Institute of Theoretical Physics, University of Warsaw,
             ul. Ho\.za 69, PL-00-681 Warsaw, Poland}

 
\date{\today}
 
\maketitle
 
 
\begin{abstract}
The nuclear quadrupole moment of the $I^\pi$=3/2$^-$ excited nuclear state 
of $^{57}$Fe at 14.41\,keV,
important in M\"{o}ssbauer spectroscopy,
 is determined from 
the  large-scale
 nuclear shell-model  calculations for $^{57}$Fe and also from
the electronic {\em ab initio} and 
density functional theory calculations
including solid state and electron correlation effects
for the molecules Fe(CO)$_5$ and Fe(C$_5$H$_5$)$_2$.
Both independent methods yield very similar results. The recommended value is 
0.16(1)\,eb. The NQM of the isomeric 10$^+$ in $^{54}$Fe has also been
calculated. The new value (0.5\,eb),
consistent with the perturbed angular distribution
data, is  by a factor of two larger than the currently 
recommended value. 
\end{abstract}
 
\pacs{21.10.Ky, 21.60.Cs, 31.15.Ar, 31.30.Gs, 
 76.80.+y, 27.40.+z}
]

\narrowtext
 
M\"{o}ssbauer spectroscopy of $^{57}$Fe plays an important role in the
structural determination of iron containing solid state compounds.
In principle, the nuclear quadrupole moment (NQM)
of the isomeric  $I$=3/2  state in $^{57}$Fe can be 
determined from  M\"{o}ssbauer data;
however, the analysis requires the calculation 
of the electric field gradient (EFG).
As these atomic calculations are quite involved, studies employing
different methods arrived at quite distinct results; the values of 
NQM in the range from --0.19 to +0.44\,eb have been reported 
\cite{rusakov}. 
This quite unsatisfying situation could  also not have been
settled by nuclear structure calculations of the NQM as calculations
within the nuclear shell model, the most reliable tool for such studies,
had to be performed in strongly truncated model spaces and with rather
untested effective interactions. In recent years, decisive progress has
been achieved in both the atomic calculations of the EFG and in nuclear
shell model studies.

In 1995, Dufek and co-workers  applied the density functional 
theory for a series of iron-containing solid state compounds. For $^{57}$Fe
they obtained a 
 NQM of 0.16 eb \cite{schwarz1},  in contradiction with the previously 
accepted value of 0.082\,eb obtained from Hartree-Fock (HF) EFG calculations 
\cite{duff} and from truncated
nuclear shell-model  calculations combined with the perturbed angular
distribution data
\cite{vajda}. In  subsequent work, Su and Coppens obtained a  NQM 
value of 0.12(3)\,eb using Sternheimer-corrected EFGs \cite{su}.
In this Letter, we shall demonstrate that 
state-of-the-art nuclear and atomic physics 
calculations  lead to the same NQM for
the $^{57}$Fe isomeric state, settling a long-standing controversial
issue.

We shall begin with the nuclear physics discussion.
The nucleus $^{57}$Fe has two low-lying $3/2^-$ states which are
experimentally split by only 353 keV. To describe the structure of such
nearly degenerate states is quite demanding. Clearly, the large-scale shell model
is the method of choice. 
Due to recent progress in programming and hardware development,
modern shell-model calculations based on
microscopic
effective interactions can handle configuration spaces that were 
prohibitively large only  several years ago  \cite{SMrev}. More specifically,
modern diagonalization shell-model codes  can now
handle medium-mass nuclei ($A$=50-60) in the middle of the $pf$ shell {\em in  full
$0\hbar\omega$ space}. To put things in perspective, the shell-model calculations
of the Utrecht Group \cite{vennink}, which were used in Ref.~\cite{vajda}
to extract the NQM of $^{57}$Fe, 
restricted the number of  holes in the $f_{7/2}$ 
orbit to three.

Shell-model calculations depend crucially on
two factors: the model space and the effective interaction. 
Our calculations have been
performed using the code NATHAN 
\cite{Caurier99}. NATHAN has been developed in the
$jj$-coupling scheme using quasispin formalism. We adopt a version of the
code adapted to shared-memory parallel machines.  
For the two lowest $3/2^-$ levels, we assumed the
 truncated  space (containing 8,120,105 $I^\pi$=3/2$^-$ states)
in which maximally 6 nucleons were allowed to
be excited from the $f_{7/2}$ orbital to the rest of the $pf$ shell. 
To test the convergence of our results, we have performed a full pf shell
calculation for the lowest
$3/2^-$ state.
This calculation includes 
25,743,302  states ($\sim$2$\cdot$10$^{13}$ non-zero
matrix elements)  and is one of 
the largest shell-model diagonalization  performed to date. 
For all adopted interactions, the results of the truncated and complete
calculations are identical.

During the last few 
years, a considerable effort  went into the development of  effective
interactions in the $pf$ shell. In this work
three  different effective  interactions have been employed: KB3F, KB3G, and FDP6.
The interactions KB3F \cite{KB3F}
and KB3G \cite{KB3G} are both reasonable attempts to correct the defects
of the well-known KB3 interaction \cite{KB3} in the upper part of the $pf$-shell.
They differ slightly in their collectivity of states around $^{58}$Ni.
As far as $^{57}$Fe is concerned,
  KB3G is clearly the interaction of choice;
it has been shown to be  very successful in 
describing  experimental data (including energy levels and electromagnetic
properties)   in the mass region
$A$=50-52 and around $A$=56 \cite{KB3G}.
The FDP6 interaction \cite{FDP6} was  originally fitted to the spectroscopic
properties of $f_{7/2}$ nuclei and has since  been extended to nuclei
around the $N$=$Z$=28 shell closure \cite{Otsuka}. For the effective charges, we took
the quadrupole charges
1.5$e$ for
protons and 0.5$e$ for neutrons when calculating  E2 transitions and
moments,  and the spin and orbital gyromagnetic factors
$g^s$=0.75$g^s_{\rm bare}$,
$g^l_\pi$=1.1\,$\mu_N$,
and $g^l_\nu$=$-$0.1\,$\mu_N$ for M1 transitions and moments, e.g. \cite{Bohr}.
We adopt the oscillator parameter $b=1.01 A^{1/6}$ fm.

Our shell-model calculations reproduce well the experimental level scheme of $^{57}$Fe.
In particular, the low-lying $I^\pi$=3/2$^-$ doublet  is predicted  by theory. Experimentally,
the energy 
difference between these states is 353 keV,  while theoretically it is
526, 209, and 465\,keV using the KB3F, KB3G, and FPD6 interactions, 
respectively. The third 3/2$^-$
state appears much higher in energy, at about 1.6\,MeV. 
Table~\ref{smdiab} summarizes our shell-model results
 for the quadrupole 
and magnetic moments of the two lowest $3/2^-$ states
in $^{57}$Fe. It is seen that these close-lying states in the doublet
are predicted to have
very different shapes and magnetic moments, i.e., their  E2 and M1 moments
have different signs. Consequently, even small changes in the shell-model
interaction, hence in the coupling between these states, significantly impact
theoretical predictions.
We note that
a convergence of the shell-model results 
is not reached if only three $f_{7/2}$ holes are allowed,
as was done in the previous truncated calculations \cite{vennink}.
For example,
the value of the NQM of the 3/2$^-_1$ state,  $Q(3/2^-_1)$, obtained in such
truncated calculations with the KB3G 
interaction, is 0.097\,eb, as compared to the exact value of
0.064\,eb.

 In order to assess the sensitivity of the results on the adopted
effective force, 
we computed overlaps between wave functions of the $3/2^-$ states
obtained with the different interactions. It turned out that the spaces of the
two lowest $3/2^-$ states in KB3F and KB3G are practically (up to 95\%)
the same. That is, 
both  $3/2^-$ wave functions obtained with KB3F can be approximately
derived from the $3/2^-$ doublet calculated with KB3G by simple rotation.
 This does not hold
for KB3G and FDP6. Since  FPD6 tends to excite particles to
the $f_{5/2}$ shell rather than  the $p_{3/2}$ shell
(FPD6 puts the $f_{5/2}$
orbit too low at $N$=28),
the contributions from higher-lying $3/2^-$ states
amount to about 37\%.

Due to the great sensitivity  of  the shell-model predictions caused by 
the not-very-well-controlled 
off-diagonal coupling between the $3/2^-$ doublet, the calculations need to
be constrained by the available experimental data. To this end, we choose
the magnetic moment of the $3/2^-_1$ state,
which has been precisely determined experimentally,
$\mu(3/2^-_1)$=$-$0.1549(2)\,$\mu_N$ Assuming that the ``true" wave functions of
the $3/2^-$ doublet are given by a simple rotation of the shell-model states,
\begin{equation}\label{mixing}
|3/2^-\rangle = \alpha_1  |3/2^-_1;{\rm SM}\rangle  + \alpha_2  |3/2^-_2;{\rm SM}\rangle,
\end{equation} 
one can determine the mixing amplitudes $\alpha_1$ and $\alpha_2$ 
($\alpha_1^2+\alpha^2_2=1$)
by requesting that the measured value of $\mu(3/2^-_1)$
be  reproduced. The results of such  two-level mixing calculations are displayed
in Table~\ref{smadiab}. Contrary to
the  pure shell-model results, one obtains a 
satisfying agreement between KB3F and KB3G, and the data. In particular,
the quadrupole moments predicted
by these interactions are  very similar, $Q(3/2^-_1)$$\approx$0.17\,eb. The only serious
difference is the sign of the magnetic moment of the $3/2^-_2$ state. Unfortunately,
this quantity is poorly determined
experimentally: $\mu(3/2^-_2)$$<$0.6\,$\mu_N$ The predictions
of FPD6 for the E2 moments are rather far from the data: 
both $3/2^-$ states are calculated
to be practically spherical,
  and  the $B(E2)$ transition connecting these states
is enhanced  by a factor of $\sim$40.
 Part of this failure comes from
  too-low a placement of the $f_{5/2}$ level in FPD6, as discussed
 in detail in Ref.~\cite{KB3G},  where more examples of KB3G and FPD6 
calculations can be found. Therefore,
we conclude that the 
best shell-model estimate 
of the quadrupole moment of the  $3/2^-_1$ state in $^{57}$Fe  is 0.16\,eb.

In Ref.~\cite{Sprouse} it has been pointed out that the ratio of quadrupole
moments of the isomeric $10^+$ state in $^{54}$Fe to the $3/2^-$ state
in $^{57}$Fe, 
$Q[^{54}$Fe$(10^+)]$/$Q[^{57}$Fe$(3/2_1^-)]$,
 represents a stringent constraint. 
Possible uncertainties
in the atomic EFG calculations cancel largely in
the experimental determination of
the ratio from M\"ossbauer data. The most recent value is
$Q[^{54}$Fe$(10^+)]$/$Q[^{57}$Fe$(3/2_1^-)]$ =3.62$\pm$ 0.22 \cite{Hass}.
Unfortunately, the NQM of the isomeric state in $^{54}$Fe is not known
experimentally. We have thus calculated the NQM
of the isomeric $10^+$ state in $^{54}$Fe in a large-scale shell-model
study
in which we
allowed 8 particles to be promoted from the $f_{7/2}$ orbital to the
rest of the shell. We have checked the convergence of our results by
performing calculations in which only 6 particles could be promoted; the
NQMs were the same at both levels of truncation. For all the interactions
used, we find rather similar NQMs: 0.51 eb (KB3F), 0.50 eb (KB3G), and 0.56 eb (FDP6).
Using the NQMs of Table~\ref{smadiab} for the isomeric state in $^{57}$Fe, these
values correspond to 
$Q[^{54}$Fe$(10^+)]$/$Q[^{57}$Fe$(3/2_1^-)]$ 
ratios which agree rather well with the measurement
for the KB3F and KB3G interactions 
(2.9 and 3.1, respectively). For the FDP6 interaction, the 
ratio is significantly too large, as the NQM of the isomer in $^{57}$Fe
is severely underestimated. We note that our
predicted values  for the magnetic moment of the $10^+$ state in
$^{54}$Fe (6.4\,$\mu_N$ in KB3F, 6.5\,$\mu_N$ in KB3G,  and 7.1\,$\mu_N$
in FDP6) compare well with the experimental value of 7.28$\pm0.01$
n.m \cite{Rafailovich}.

We now turn to our atomic physics studies. 
The use of the   density functional theory  for the calculation of EFGs in 
transition metals  is questionable. 
Recent calculations on CuF showed a variation 
of the Cu EFG ranging from +0.50 au for the local density approximation (LDA), and
+0.44 au at the generalized gradient approximation (GGA) level (BPW91)
  to +0.07 au at 
the hybrid level (B3PW91) \cite{schwerd,lenthe}, as compared to the 
experimental value of $-$0.31(2) au \cite{hoeft}. In contrast, relativistic ab initio 
coupled-cluster calculations give $-$0.34 au  \cite{schwerd}, in perfect agreement with the 
experimental result.

   In this work, 
we carried out density functional as well as ab initio 
calculations for the molecules Fe(CO)$_5$ and Fe(C$_5$H$_5)_2$.
We have adopted
a wide range of exchange and correlation functionals 
for the electronic structure calculations of the free molecules
Fe(CO)$_5$ and Fe(C$_5$H$_5)_2$ (for the terminology see Refs. 
\cite{schwerd,book}): Hartree-Fock-Slater 
(HFS), X$\alpha$, Local Density Approximation (LDA), 
the GGA functionals B-HFS (Slater exchange plus Becke nonlocal exchange), 
B-LYP (B-HFS plus the correlation functional of Lee, Yang and Parr), B-PW91 
(B-HFS plus the correlation functional of Perdew and Wang), the hybrid GGA 
functionals B3-LYP (Becke three-parameter functional), B3-PW91 
(same as B3-LYP except with the nonlocal correlation term of Perdew and Wang), 
BHH (Becke half-and-half together with the LYP correlation functional), and 
BHH-LYP (same as BHH but with
0.5 of the Becke nonlocal exchange term added to 
the energy). For comparison with DFT, we carried out the  HF 
many-body perturbation theory (MBPTn) up to order n=4, as well as 
coupled-cluster singles doubles including non-iterative triple excitations 
[CCSD(T)] \cite{frisch}. 
The electronic coupled cluster calculations required three months of
CPU time and 20 Gbytes of disk storage on a Origin 2000 SGI.
For Fe(CO)$_5$ we investigated solid state effects 
to the iron electric field gradient  by performing HF, B3LYP, and LDA 
calculations using the program CRYSTAL98 \cite{crystal} and the solid state 
structure given by B\"{o}se and Bl\"{a}ser \cite{boese}. Detailed structural 
data and basis sets used will be published elsewhere \cite{schwerd2}.

   The results of our electronic structure calculations for the iron EFGs of Fe(CO)$_5$ and 
Fe(C$_5$H$_5)_2$ are shown in Table~\ref{atomtable}.
 We first note that the 
single-reference
many-body perturbation theory shows extreme oscillatory behavior and is
practically useless for the determination of transition element EFGs. DFT  results
range from 1.09\,au to 1.57\,au for Fe(CO)$_5$, and from 
1.36\,au to 2.49\,au for Fe(C$_5$H$_5)_2$,
depending on the functional applied. If we accept the coupled-cluster
result as the most accurate value,   the hybrid GGA functionals perform
well for Fe(CO)$_5$, while  the non-hybrid GGA functionals are preferred for
Fe(C$_5$H$_5)_2$. 
We note that for CuCl the BHH functionals gave the best description  \cite{schwerd}. 
This is clearly not a satisfying situation. On the other hand, an 
encouraging result  is that 
solid state effects from nearest-neighbor interactions in Fe(CO)$_5$ are small and 
can basically be neglected. Based on  the CCSD(T) EFGs,  we obtain from the
M\"{o}ssbauer data of Fe(CO)$_5$) (2.57 mm/sec \cite{greenwood}) and Fe(C$_5$H$_5)_2$ 
(2.4 mm/sec \cite{collins}) a NQM of 0.177\,eb  and 0.159\,eb,  respectively.
However, basis set incompleteness and relativistic effects may increase the iron 
EFGs and, therefore,  further decrease the NQM \cite{schwerd2}. Consequently,
our best estimate 
using EFGs together with M\"{o}ssbauer data for these molecules is 0.15(2)\,eb. In
order to improve further on these results, large-scale relativistic coupled-cluster
calculations are necessary, which are currently not feasible for such big
molecules. We emphasize,
however, that EFGs obtained from current DFT for transition
element compounds should be taken with some care as the results
in Table~\ref{atomtable} show.

In summary, the quadrupole moment of  the first $3/2^-$ state in $^{57}$Fe 
at 14.41\,keV,
important in M\"{o}ssbauer spectroscopy, 
has been calculated using  state-of-the-art nuclear and 
atomic  models. Both calculations yield results which are consistent with
$Q(3/2^-_1)$$\approx$0.16\,eb, in  nice agreement with the recently reported
value \cite{schwarz1} and the results of 
Ref.~\cite{Fan}. As a by-product of this work,
the NQM of the isomeric 10$^+$ in $^{54}$Fe has also been 
calculated. The new value, consistent with the perturbed angular distribution
data, is 0.5\,eb, i.e.,  a factor of two greater than the currently 
recommended value \cite{Hass}.

This research was supported by the U.S. Department of Energy under
Contract Nos.\ DE-FG02-96ER40963 (University of Tennessee),
DE-FG02-97ER41019 (University of North Carolina),
DE-AC05-00OR22725 with UT-Battelle, LLC (Oak Ridge National
Laboratory), DE-FG02-92ER40694 (Tennessee Technological University),
the Danish Research Foundation,
the Schweizer Nationalfond (grant 20-61822.00),
the Marsden Fund Wellington, the Royal Society of New Zealand, and the
Auckland University Research Committee. 
The Auckland group is indebted to the Alexander von Humboldt
Foundation for financial support. GMP wants to thank the Carlsberg
Foundation for a fellowship.

\newpage

 

\begin{table}[ht]
\caption{
Results of the shell-model calculations for the quadrupole 
and magnetic moments of the  two lowest $3/2^-$ states
in $^{57}$Fe using  effective interactions
KB3F, KB3G, and FPD6. The experimental value
of $\mu(3/2^-_1)$ is --0.1549(2) $\mu_N$ \protect\cite{dane}.
}\label{smdiab}
\begin{tabular}{c|rrrr}
& & KB3F & KB3G & FPD6  \\
 \hline
 $Q$ & $3/2^-_1$ & 0.16 & 0.06 & $-$0.17 \\
  (eb) &  $3/2^-_2 $ & $-$0.16 & $-$0.07 & 0.17 \\
  \hline
   $\mu$ & $3/2^-_1$ & $-$0.32 & $-$0.49 & $-$0.51 \\
    ($\mu_N$) & $3/2^-_2$ & 0.26 & 0.23 &  0.49 
\end{tabular}
\end{table}

 \begin{table}[ht]
\caption{
Results of the two-level mixing calculations
(mixing amplitudes, quadrupole and magnetic moments,
and transition probabilities)
 for the  two lowest $3/2^-$ states
in $^{57}$Fe using  effective interactions
KB3F, KB3G, and FPD6.  The mixing amplitudes $\alpha_1$ and $\alpha_2$ 
(\protect\ref{mixing})
for the $3/2^-_1$ state
have been adjusted to reproduce the experimental value of
$\mu(3/2^-_1)$=$-$0.155 $\mu_N$ 
$B(M1)$ and $B(E2)$ denote $3/2^-_2\rightarrow 3/2^-_1$ transition rates.
See text for details.
}\label{smadiab}
\begin{tabular}{c|cccccc}
  & $\alpha_1$ & $\mu(3/2^-_2)$ &  $Q(3/2^-_1)$ &  $Q(3/2^-_2)$ & 
$B(M1)$  & $B(E2)$  \\
 & &  ($\mu_N$) & (eb) & (eb) & ($\mu_N^2$) & (e$^2$fm$^4$) \\ 
 \hline
KB3F & 0.99 & 0.10 & 0.18 & $-$0.12 & 0.13 & 4 \\
KB3G & 0.92 & $-$0.10 & 0.16 & $-$0.17 & 0.09 & 25 \\
FDP6 & 0.87 &     0.14 & 0.02 & $-$0.02 & 0.10 & 228 \\
{\bf exp.}\protect\cite{dane} &             & $<$0.6 &  &   & 0.07  &   5  
\end{tabular}
\end{table}

\begin{table}[ht]
\caption{Calculated iron electric field gradient for Fe(CO)$_5$ and Fe(C$_5$H$_5)_2$
at various levels of theory. SS denotes solid state calculations.
All values are in au.}\label{atomtable}
\begin{tabular}{c|cc}
Method & Fe(CO)$_5$ & Fe(C$_5$H$_5)_2$ \\
\tableline\\
 & \multicolumn{2}{c}{\it{ab-inito}}  \\
HF & 1.394 & 3.157 \\
SSHF & 1.367 & - \\
MBPT2 & 3.511 & 1.563 \\
MBPT3 & -0.740 & 3.385 \\
MBPT4 & 9.484 & -0.250 \\
CCSD(T) & 1.394 & 1.496 \\
\\
 & \multicolumn{2}{c}{\it{DFT}}  \\
X$\alpha$ & 1.148 & 1.374 \\
HFS & 1.092 & 1.363 \\
LDA & 1.154 & 1.359 \\
SSLDA & 1.122 & - \\
BHFS & 1.187 & 1.434 \\
BLYP & 1.203 & 1.425 \\
BPW91 & 1.203 & 1.377 \\
B3LYP & 1.393 & 1.854 \\
SSB3LYP & 1.357 & - \\
B3PW91 & 1.395 & 1.806 \\
BHH & 1.547 & 2.429 \\
BHHLYP & 1.573 & 2.488 
\end{tabular}
\end{table}


\newpage

\begin{thebibliography}{99}
 
\bibitem{rusakov} V.S. Rusakov and D.A. Khramov, Bull. Russ. Acad. Sci. 
Phys. {\bf 56}, 1118 (1992).
 
\bibitem{schwarz1} P. Dufek, P. Blaha, and K. Schwarz, Phys. Rev. Lett. {\bf 75}, 
3545 (1995).

\bibitem{duff} K.J. Duff, K.C. Mishra, and T.P. Das, Phys. Rev. Lett. {\bf 46}, 1611
(1981).
 
\bibitem{vajda} S. Vajda, G.D. Sprouse, M.H. Rafailovich and J.W. No\'e, 
Phys. Rev. Lett. {\bf 47}, 1230 (1981).

\bibitem{su} Z. Su and P. Coppens, Acta Cryst. A {\bf 52}, 748 (1996). 


\bibitem{SMrev} K. Langanke and A. Poves, Nucl. Phys. News {\bf 10}, 3 (2000).

\bibitem{vennink} R. Vennink and P.W.M. Glaudemans, Z. Phys. A {\bf 294}, 241 (1980).
 
\bibitem{Caurier99} E. Caurier, G. Mart\'{\i}nez-Pinedo, F. Nowacki, 
A. Poves, J. Retamosa, and   A.P. Zuker, 
Phys. Rev. C {\bf 59}, 2033 (1999).  

\bibitem{KB3F} H. Kaiser {\it et al.}, Nucl. Phys. {\bf A669}, 368
(2000).

\bibitem{KB3G} A. Poves, J. Sanchez-Solano, E. Caurier, and F. Nowacki, 
nucl-th/0012077; to be published.

\bibitem{KB3} G. Mart\'{\i}nez-Pinedo, 
 A.P. Zuker, A. Poves, and E. Caurier,
Phys. Rev. C {\bf 55}, 187
(1997).


\bibitem{FDP6} W.A. Richter, M.G. Vandermerve, R.E. Julies, and B.A.
Brown, Nucl. Phys. {\bf A523}, 325 (1991).

\bibitem{Otsuka} T. Otsuka, M. Honma, and T. Mizusaki,
 Phys. Rev. Lett.  {\bf 82}, 1588 (1998).


\bibitem{Bohr} A. Bohr and B.R. Mottelson, {\it Nuclear Structure},
 Vol. II (Benjamin, New York, 1975).


\bibitem{dane} M.R. Bhat, Nucl. Data Sheets {\bf 85}, 415 (1998).

\bibitem{Sprouse} S. Vajda, G.D. Sprouse, M.H. Rafailovich, and J.W. Noe,
Phys. Rev. Lett. {\bf 47}, 1230 (1981).

\bibitem{Hass} M. Hass {\it et al.}, Nucl. Phys. {\bf A414}, 316 (1984).

\bibitem{Rafailovich} M.H. Rafailovich, E. Dafni, J.M. Brennan and G.D.
Sprouse, Phys. Rev. C {\bf 27}, 602 (1983).

 
\bibitem{schwerd} P.Schwerdtfeger, J. K. Laerdahl, and M. Pernpointner, 
J. Chem. Phys {\bf 111}, 3357 (1999).

\bibitem{lenthe} E. van Lenthe and E. J. Baerends, J. Chem. Phys {\bf 112}, 
8279 (2000).
 
\bibitem{hoeft} J. Hoeft, F.J. Lovas, E. Tiemann,  and T. T{\o}rring, 
Z. Naturforsch {\bf 26a}, 240 (1970).
 
\bibitem{book}{\it Recent Developments and Applications of Modern Density Functional 
Theory, Theoretical and Computational Chemistry}, ed. by J.M. Seminario 
(Elsevier, Amsterdam, 1996). 
 
\bibitem{frisch} M.J. Frisch et al., program Gaussian 98, Gaussian Inc., 
Pittsburgh PA, 1998.
 
\bibitem{crystal} V.R. Saunders et al., program CRYSTAL98, 
University of Torino, Torino (1998).

\bibitem{boese} R. B\"{o}se and D. Bl\"{a}ser, Z. Krist. {\bf 193}, 289 (1990). 

\bibitem{schwerd2} P. Schwerdtfeger, T. S\"{o}hnel, M. Pernpointner, and
J. K. Laerdahl,  to be
published.
 
\bibitem{greenwood} R. Greatrex and N.N. Greenwood, Discuss. 
Faraday Soc. {\bf 47}, 126 (1969).
 
\bibitem{collins} R.L. Collins, J. Chem. Phys. {\bf 42}, 1072 (1964).

\bibitem{Fan} M. Fanciulli {\it et al.}, Phys. Rev. B {\bf 59}, 3675 (1999).
 
\end{thebibliography}
\end{document}